\documentclass{ws-p8-50x6-00}
\def\rightnote{Statistics of Vacua\hfill
String Phenomenology 2003, Durham}

\begin{document}

\title{Statistics of String vacua}

\author{Michael R. Douglas}
\address{Department of Physics and Astronomy, Rutgers University\\
Piscataway, NJ 08855-0849 USA\\
{\it and}\\
I.H.E.S., Le Bois-Marie, Bures-sur-Yvette, 91440 France}


\maketitle

\newcommand{\adref}{0307049}
\abstracts{
We give an introduction to the statistical approach to studying
vacua of string/M theory,
and discuss recent results of 
Ashok and Douglas on
counting supersymmetric flux vacua in type IIb
Calabi-Yau compactification.\\
\noindent
To appear in the proceedings of the 2003 String Phenomenology
workshop in Durham, UK.}

% statistics talk
\newcommand{\page}{}
\newcommand{\p}{}
\newcommand{\NN}{{\mathcal \bf N}}
\hyphenation{super-symmetric super-symmetry}
\hyphenation{non-super-symm-etric di-men-sion-al}
\def\blue{}
\def\red{}
\newcommand{\sfont}[1]{{\tiny #1}}
\newcommand{\sref}[1]{{\blue (#1)}}
\newcommand{\heading}[1]{\begin{center}\large\bf #1\end{center}}
\newfam\black
\font\blackboard=msbm10 %scaled \magstep1
\font\blackboards=msbm7
\font\blackboardss=msbm5
\textfont\black=\blackboard
\scriptfont\black=\blackboards
\scriptscriptfont\black=\blackboardss
\def\Bbb#1{{\fam\black\relax#1}}
\def\sgn{{\rm sgn\ }}
\def\etal{{\it et.al.}}
\def\slashslash{{/\hskip-0.2em/}}
\def\CA{{\cal A}}
\def\CC{{\cal C}}
\def\CD{{\cal D}}
\def\CF{{\cal F}}
\def\CH{{\cal H}}
\def\CI{{\cal I}}
\def\CJ{{\cal J}}
\def\CL{{\cal L}}
\def\CM{{\cal M}}
\def\CN{{\cal N}}
\def\CO{{\cal O}}
\def\CP{{\cal P}}
\def\CT{{\cal T}}
\def\CW{{\cal W}}
\def\BC{\Bbb{C}}
\def\BH{\Bbb{H}}
\def\BM{\Bbb{M}}
\def\BP{\Bbb{P}}
\def\BR{\Bbb{R}}
\def\BX{\Bbb{X}}
\def\BZ{\Bbb{Z}}
\def\mapr{\mathop{\longrightarrow}\limits}
\def\half{{1\over 2}}
\def\GeV{~{\rm GeV}}
\def\TeV{~{\rm TeV}}
\def\Coh{{\rm Coh}~}
\def\Cohc{{\rm Coh}_c~}
\def\Mod{{\rm Mod}~}
\def\ind{{\rm ind}~}
\def\Tr{{\rm Tr}~}
\def\tr{{\rm tr}~}
\def\grad{{\rm grad}~}
\def\CY#1{CY$_#1$}
\def\rk{{\rm rk}~}
\def\Im{{\rm Im}~}
\def\Hom{{\rm Hom}}
\def\Ext{{\rm Ext}}
\def\Vol{{\rm Vol~}}
\def\Stab{{\rm Stab}}
\def\End{{\rm End~}}
\def\ib{{\bar i}}
\def\jb{{\bar j}}
\def\zb{{\bar z}}
\def\pp{\partial}
\def\pb{{\bar\partial}}
\def\I{{I}}
\def\II{{II}}
\def\IIa{{IIa}}
\def\IIb{{IIb}}
\def\vev#1{{\langle#1\rangle}}
\def\vvev#1{{\langle\langle#1\rangle\rangle}}
\def\ket#1{{|#1\rangle}}
\def\Dslash{\rlap{\hskip0.2em/}D}
\def\dual{{v}} % or check mark, or whatever
\def\intersect{\cdot}
\def\Nbar{{\bar N}}
\def\NRR{{N_{RR}}}
\def\NNS{{N_{NS}}}
\def\bigvev#1{\bigg\langle{#1}\bigg\rangle}

\section{Introduction}

What is the most important problem of string phenomenology?

Most would probably say that it is to find a compactification which
reproduces the Standard Model in all its details, and makes new
predictions.  More realistically, one could settle for qualitative
agreement with the Standard Model, as long as one gets new
predictions.  

If one interprets ``qualitative'' loosely enough, this has been done,
but the standard which has been met is not very high.  While the
general structure of supersymmetric grand unification was already realized
in the early works, many of the important features, especially
the hierarchy of scales and the small cosmological constant, 
not to mention specific parameter values,
were controlled by non-perturbative effects, which were not at all understood
at that time.  Besides these classic problems,
string and M theory constructions come with new problems
of their own, which require solution or explanation: why our universe
appears four-dimensional, what chooses a particular compactification
or other extra dimensional structure, and so forth.

Building on the understanding of non-perturbative string/M theory 
achieved in the mid-late 1990's, it is now possible to study
compactifications which address these goals.  Still, no
model has been proposed which meets all of the requirements.
Should we be worried?\p

One can argue that, to the contrary, this is {\bf good} news for any
hope for making predictions from string theory.  The argument is
simply that we have only explored a tiny fraction of the
compactifications now believed to exist, and these seem to be a fairly
random cross section chosen not because they are more likely to work,
but more because of ease of analysis and historical accident.  \p

If we had already found satisfactory candidates in this small sample,
it is likely that the full set would contain large numbers of theories
which agree with the Standard Model {\bf in every detail}.  Even worse,
different theories out of this set might 
lead to very different predictions.  In a worst-case scenario, 
string theory might not be testable at all.

To bring home this point, let us look at the variety of string/M
theory constructions, and ask how many might be candidates. 
Focus on Calabi-Yau compactification, where some numbers
are known.

The number of distinct CY threefolds is believed to be $10^5$--$10^6$.
All these constructions involve additional choices -- a choice of
bundle or brane, with comparable multiplicity, which directly affect
the spectrum.  Choices of flux or nonperturbative gauge theory vacuum
probably bring in much larger multiplicities -- numbers 
$\sim 10^{100}$ are often cited (more  below). 

In a weakly coupled construction, it is not hard to exclude models with
exotic matter of various types.  But most constructions contain sectors
with $O(1)$ couplings.  Many proposals exist for strongly coupled
sectors (composite models, supersymmetric technicolor, the supersymmetry
breaking sector, or just hidden sectors) and our current understanding 
of nonperturbative gauge theory tends to support these claims.
Thus, one cannot throw out such models from the start.  On the positive
side, one is better able to check if a specific model works.

Even leaving out considerations of moduli fixing and couplings,
numbers like $10^{10}$ qualitatively distinct models seem very
plausible.  So far, the number which have been considered in any
detail is more like $100$.

If one agrees with this argument, then one is led to the belief that
constructing a model which agrees with the Standard Model in
detail is {\bf not} the primary goal we should be pursuing
at this point:
\begin{itemize}
\item If it is possible within our present limitations,
this is a negative result.
\item If it is not possible within our present limitations,
one cannot get a result.
\end{itemize}
While this conundrum oversimplifies the situation, it deserves
consideration.  \p

Of course there are ways around it.  Before focusing on one, 
let us at least mention the two others we know of.\p

One is to look for predictions which cannot be matched by four
dimensional effective field theory, for example short range
modifications to gravity.  This is important, and many speakers here
are discussing it.  On the other hand, we have no clear reason to
think such effects must exist. Indeed, the ``traditional''
Kaluza-Klein scenario still seems to fit evidence such as unification
of coupling constants better than any of the more recently proposed
variations, and must be taken seriously.

The other is to look for some {\it a priori} principle which selects
among the possible vacuum configurations.  Given such a principle, it
is of course much more interesting to know if the vacuum it selects
can reproduce the observations. 

As an example to illustrate ``Vacuum Selection Principles,'' let me
give you my best idea along these lines.
\p
One might imagine  (see for example \cite{dns})
that our vacuum is in some sense ``the most symmetric''
among the various possibilities.
Besides esthetics, many other candidate
principles -- for example, maximizing some natural wave function of the
universe -- seem likely
to prefer such vacua.
\p

On the face of it, this proposal is absurd.  Higher dimensional models
with more supersymmetry are obviously more symmetric.  
By rights, the ``most symmetric'' four dimensional theory, is the one with
the most gauge symmetry.  F theory examples are known with
rank $10^5$ gauge groups.
\cite{Canska}
The Standard Model is not even in the running.
\p

However, one might imagine that there exists some unstable
nonsupersymmetric vacuum with a huge gauge group, which is preferred
by Planck scale cosmology.  It would then roll down to our physical
vacuum.  Then, the ``right vacuum'' would be near this preferred
symmetric point.

This is as close to a plausible {\it a priori} principle as I have
come, but using it still requires some fairly detailed knowledge about
the set of possible vacua, and the configuration space which contains
the vacua.  Most such ideas require even more information.
It does not seem reasonable to hope for a principle which will tell
one in advance which string theory, Calabi-Yau, brane configuration
etc. to look at.  \p

Of course, it could be that the tests we already know of (or will know
in ten years time) are already stringent enough to select out zero or one
vacua.  Maybe if we could work with them, an {\it a priori} principle
would not be necessary. \p

And, we should keep in mind, that there is no guarantee that any
{\it a priori} vacuum selection principle exists.  We only have one
sample, and the question of why we observe this one need not have any
better answer than ``because we are here.'' \p

This is not
an anthropic argument, which is a much more specific and 
predictive type of argument.  It is just a conservative interpretation
of the goal of science: to explain what we see, not why
we see it.

\section{Statistics of vacua}

So what to do?  There is a third approach, which has been advocated for
some time by Dine \cite{dine}; it is to look for ``generic'' predictions of
string theory.  Can we make this idea precise?

Any attempt to make generic predictions has to be founded on the idea
that we know ``all'' string vacua or at least some representative set,
in the sense that the distribution of a property of interest is the
same in the representative set as in the whole.  At present we can
only hypothesize that the sets of vacua we can study concretely could
be representative, and eventually check this by the results.  For
example, one can argue that the set of all type \II\ compactifications
on Calabi-Yau, to the extent that it can be studied at arbitrary
volume and string coupling, could be a large fraction of the vacua,
because the other known large classes of vacua are believed to be 
dual to these.

This at least gets us started, but it is clear that any representative
class of vacua is far too complicated to study in any detail at
present.  We need to ask simpler questions, which might give us some
picture of this huge ``landscape'' of theories.

To do this, we proposed in \cite{mrdstat} to work as follows.  The
idea is to make a precise hypothesis for an approximate description of
this set: the vacua are vacua of a specific {\blue ensemble of
effective field theories}, something like a list or set of theories
which we believe can come out of string/M theory.  While the ensemble
should be precisely specified, we need not claim that it exactly
represents the set of string/M theory vacua, only that it represents
it well enough for our purposes.\p

We then can proceed in two directions:
\begin{itemize}
\item We can test whether our hypothesized ensemble is accurate, by
comparing with actual string/M theory constructions.
\item We can find out what fraction of vacua out of our ensemble
meet a specified phenomenological test.
\end{itemize}

\page

Let us give a very simple example to illustrate the point, by asking
the question:
Out of {\bf all} the four dimensional vacua obtained by string/M
theory compactification, {\bf how many} of them are effective field
theories with $SU(3)\times SU(2)\times U(1)$ gauge symmetry unbroken
at low energy?  \p
If we define our terms, and if string/M theory has a precise
definition, {\bf and} if there are finitely many physically distinct
vacua, then this question has a definite answer.  \p

One can just as easily generalize the question to, out of all vacua,
how many have low energy gauge group $G$ ?  Let us denote this number
by the function
$$
d\mu[G] .
$$
While finding this function exactly is hard, perhaps
it can be approximated in some simple and useful way.
\p

For example, could it be that the rank $r={\rm rk}\ G$
of the unbroken gauge group, roughly satisfies a power law distribution,
$$
d\mu[r] \sim N \times r^{-\alpha} .
$$
If so, and if we could estimate $N$ and $\alpha$,
we could get a rough estimate for how many vacua have a rank
4 gauge group, without much effort.  One could go on to study
the distribution $d\mu[N_1,N_2,\ldots]$ of the ranks of the simple
factors, etc.
Although this may sound ambitious, given that the function $d\mu[r]$ is
well defined, why shouldn't it have a simple approximate description ?

\subsection{Distribution of quiver gauge theories}

Let me turn to another problem along these lines, for which I can even
suggest a simple approximate description (details can be found in
\cite{mrdstat}).  We consider $U(N)$ quiver gauge theories, {\it i.e.}
with gauge group $\prod_i U(N_i)$, and a spectrum of purely
bifundamental matter.  Such theories arise on the world-volumes of
D-branes in type \II\ string theory, embedded in the $3+1$ observable
dimensions. and constructing these is the first step in making ``brane
world'' realizations of the Standard Model, as discussed in many talks
here.

It would be quite interesting to know the distribution of gauge groups
and matter content for the theories which come out of string
compactification.  Let us focus on part of this information: the
difference between the number of multiplets $(\bar N_i,N_j)$, and the
number $(\bar N_j,N_i)$.  For a theory with $K$ factors in the gauge
group, which would arise by wrapping branes on $K$ distinct cycles,
these are $K(K-1)/2$ {\it a priori} independent numbers;
we can summarize them in an
antisymmetric matrix, the ``intersection matrix'' $I_{ij}$.

These numbers counts chiral matter multiplets which cannot be lifted
by mass terms, and thus generalize the ``number of generations'' in
the Standard Model.  In explicit brane constructions, they are
entirely determined by the topology of the branes $B_i$ and $B_j$
carrying the $U(N_i)$ and $U(N_j)$ gauge groups.  The simplest example
of this is to consider D$6$-branes in Calabi-Yau compactification of
\IIa\ string theory: in this case, the intersection numbers are
literally topological intersection numbers between the
three-dimensional world-volumes of the branes in the CY.  Similar
formulas are known for the other types of branes, in all cases
topological.

Let us grant that the totality of type \II\ compactifications on CY,
with subsequent choices, leads to a finite set of 
vacua, each with a quiver gauge theory realized on
D-brane world-volumes.  If this set is finite, it defines a
distribution $d\mu[N_i; I_{ij}]$, the number of theories realizing
each possible choice of the $N_i$ and $I_{ij}$.
\p

In \cite{mrdstat},
we give arguments that as a matrix element $I_{ij}$ becomes large
(but not too large), this distribution goes as
\begin{equation}\label{eq:scaling}
d\mu[N_i; I_{ij}] \sim {dI_{ij}\over|I_{ij}|} ,
\end{equation}
(with a cutoff $I_{max} \sim \min(N_1,N_2)$, determined in string
theory compactification to be $I_{max} \sim 100$ by tadpole cancellation.)\p

The basic argument for this is that any given $U(N_1)\times U(N_2)$ theory
leads to a distribution $I\sim k,k^2,k^3,\ldots$
with this power-like falloff, and the ``total'' distribution obtained
by adding such distributions will also have this power-like falloff.

We then generalize to many gauge groups, by taking the
distributions of bifundamentals for each pair of gauge groups to be 
independent.  This ensemble has the great virtue of simplicity, and
should not be dismissed out of hand.  However, it is probably too simple,
as it ignores the fact that branes tend to wrap groups of cycles 
which intersect among themselves, and do not intersect between groups.

A better candidate ensemble of theories can be obtained by considering
$K\times K$ matrices which can be decomposed into blocks of size
$K_1\times K_1$, $K_2\times K_2$, and so on, and using (\ref{eq:scaling})
to describe the expected distribution of matter content in each block.
This is also very simple, and not obviously wrong.  

Of course, we are not claiming that this is an exact description of
the list of gauge theories coming out of string theory
compactification, only that it models some features of the true list.
One could go further and assert that, at the moment, this is the best
simple candidate description of the range of matter spectra coming
from brane constructions; it would be interesting to test it by
checking it against the families of models which have already been
concretely developed.  In any case, it is a precise ansatz which one
can use to study the fraction of models with a specified matter
content.

For example, one can
obtain the Standard Model by taking the gauge group
$U(3)\times U(2)\times U(1)\times U(1)$, and the
intersection matrix
$$
\left(\matrix{
0& -3& 3& 3\cr
3& 0& -1& 2\cr
-3& 1& 0& 0\cr
-3& -2& 0& 0}\right) ,
$$
and applying a subsequent orientifold projection.

In the ensemble (\ref{eq:scaling}) with $K=4$, the fraction of brane
models which realizes this spectrum is
$$
d\mu(-3)d\mu(3)d\mu(3)d\mu(-1)d\mu(2)d\mu(0) \sim 10^{-6} .
$$

If a given compactification has more than $4$ gauge groups, we need to
enumerate subsets of $4$ and compare them with the SM.  The fraction which
work depends on whether we allow exotic matter charged under the SM
gauge groups, which would live in the off-diagonal terms in the following
block decomposition:
$$
I = \left(
\matrix{ I_{SM}& I_{exotic}\cr -I_{exotic}^t& \ddots}\right)
$$
If we do not allow exotic matter, since most of the distribution has
$I_{ij}\ne 0$,
models in which the SM is realized in
a block with $K_i=4$ are very much favored.

While these ensembles are rather oversimplified, we believe that a
description of the true ensemble of brane gauge theories, which suffices
for this purpose, need not be too much more complicated.  One can refine
our estimate
by formulating more detailed ensembles, and comparing them with actual
string theory constructions.  We suspect this will
lead to similar results, say
$$
10^{-16} < {N_{SM}\over N_{{\rm all}\ G,R}} < 1 .
$$

In any case, we have formulated a quantitative sense in
which the Standard Model matter content is ``generic.''  It is not
to say that most models have this spectrum; indeed the fraction which
do is small.  But it might be large, when compared with other numbers.

\section{Flux vacua}

Perhaps the most straightforward class of vacua to which to apply these
ideas is the set of ``flux vacua'' obtained by turning on gauge field
strengths in the compact dimensions.  These have been the focus of
much recent work, in which it has been shown that their contributions
to the vacuum energy can stabilize moduli.

Flux compactifications can even realize the
observed small positive vacuum energy, and in this sense solve the
cosmological constant problem.
A particularly simple proposal of this type was made by Bousso and
Polchinski \cite{boupol} (see also Feng {\it et al} \cite{feng}).
They argued that, in compactification on a Calabi-Yau with $K$
cycles, the number of vacua with small cosmological constant could
go as $c^K$, growing exponentially with the number of cycles.
Since typical Calabi-Yau's have $K\sim 100-500$ cycles, this suggests
that string theory could have a huge multiplicity of vacua, $N_{vac}
\sim 10^{100}-10^{500}$.  Furthermore, these vacua realize a spectrum or
``discretuum'' of values for the cosmological constant, which is 
roughly uniform near zero.  Even if the distribution has no special
properties near zero, this makes it quite likely that vacua with
the small observed cosmological constant could exist just on statistical
grounds.

In more detail, let $F$ be a gauge field strength.
The equations of motion $\nabla F=0$ force it to be harmonic, so
determined by its integral over non-trivial homology cycles $\Sigma_\alpha$.
Let
$$
 N^\alpha  = \int_{\Sigma_\alpha} F
$$
be the quantized number of $F$ fluxes on the cycle $\Sigma_\alpha$,
and $K$ be the number of cycles.
\p

A qualitative description of the total energy is
is
$$
E = E_0 + {1\over l^4} \sum_{i=1}^K q_i(z)^2 N_i^2
$$
where $E_0$ is a flux-independent contribution, $q_i$ is a ``charge''
(determined by kinetic terms) and $l$ is the length scale
of the internal space.

\page
Suppose $E_0 < 0$ and $q\sim 1$, then the number of flux vacua with 
given $\Lambda = E(N)$ is roughly
\begin{eqnarray}
d\mu_{vac}(\Lambda) &\sim \int d^KN\ \delta(\Lambda-E) \\
&\sim \left(\Lambda - (E_0 l^4)\right)^{K/2-1} .
\end{eqnarray}
\p

Thus, the number of vacua with $|\Lambda| < \epsilon/l^4$ is roughly
$$
d\mu_{vac}(\Lambda\sim 0) \sim \epsilon L^{K/2-1}
$$
with $L=E_0 l^4$, substantiating the claims.

While we see the exponential emerge, and the large number of vacua,
this argument raises many questions: for example,
what determines the crucial parameters $E_0$ and $l$.

More to the point, this is only a heuristic argument, which ignored
the fact the fluxes {\bf back react} on the metric.  In the above
formula, this was expressed in the dependence $q_i(z)$ of the ``charges''
on moduli.  The actual energy is found by minimizing $z$, and this is
what determines the distribution of vacua in the moduli space.

Could it be that for many fluxes, there is no minimum apart from the
infinite volume limit?  Then, back reaction would eliminate most of
this supposed large number of vacua ?  Or, could most of them be dual
realizations of the same vacua ?

\subsection{Counting flux vacua}

In \cite{ad}, with Sujay Ashok, we answer this question by giving the
first precise estimate for the number of vacua in a family of
compactifications.  Without going into excessive detail, we work with
the \IIb\ compactifications developed by Giddings, Kachru and
Polchinski,\cite{gkp} in which the problem of finding vacua in the
full ten dimensional theory can be shown (in the large volume limit)
to precisely reduce to a problem in an $\CN=1$ effective supergravity
theory.

This effective theory has the following
chiral superfields:
$\tau=C^{(0)}+ie^{-D}$ the axion-dilaton, 
$z^i$ the complex structure moduli of $M$, 
and $\rho^i$ the K\"ahler moduli of $M$.
Their
K\"ahler potential is the same (up to truncating fields) as in
the related $\CN=2$ supersymmetric compactification with no flux,
$$
K(z,\bar z) = -\log \Im \bar z^i {\pp\CF(z)\over\pp z^i}
 -\log\Im\tau - 3\log\Im\rho .
$$

Besides a choice of CY and orientifolding, a flux compactification
sector is characterized by a quantized flux.  In \IIb\ theory these
are described by two integers for each three-cycle of $M$,
$$
N^\alpha \equiv N_{RR}^\alpha + \tau N_{NS}^\alpha =
\int_{\Sigma_\alpha} F^{(3)}_{RR} + \tau H^{(3)}_{NS} .
$$
The potential can then be computed exactly at large volume, using special
geometry and the
superpotential \cite{gvw}
\begin{eqnarray*} \label{eq:gvw}
W(z) &= &\int_M (F^{(3)}_{RR} + \tau H^{(3)}_{NS}) \wedge \Omega(z)
\equiv \int_M G \wedge \Omega(z) ; \\
\end{eqnarray*}
This $W$ and $K$ can be computed explicitly, using techniques 
developed in the study of mirror symmetry.
%\sref{Candelas, de la Ossa, Morrison, Katz, etc.}.

This superpotential depends on dilaton and complex structure moduli in
a fairly complicated way, and indeed stabilizes all of these moduli.
On the other hand, it does not depend on K\"ahler moduli, and as is
well known this $K$ exhibits ``no scale'' structure and does not
stabilize these moduli at all.  Let us completely neglect the K\"ahler
moduli for the time being and treat the others, returning to this
point below.

In this problem, it turns out that the allowed ``amount of flux,''
the number $L$ of the previous argument, is 
constrained by tadpole cancellation:
\begin{equation}\label{tadpole}
\int F \wedge H =
N(O3\ {\rm planes}) -
N(D3\ {\rm branes}) \equiv L
\end{equation}

Thus, the precise question we ask, is to count the number of
supersymmetric vacua, {\it i.e.} solutions of
$$
D_iW(z,\tau)= 0 ,
$$
satisfying this bound, as a function of $L$, up
to duality equivalences.  The duality group in this context (large
volume Calabi-Yau) is $SL(2,\BZ)$ acting on the dilaton-axion, times a
subgroup of $Sp(b_3,\BZ)$, which acts simultaneously on the choice of
flux, and on the complex structure.  A simple way to get one
representative of each duality class is to sum over all choices of
flux, but only count vacua which are stabilized at $(z,\tau)$
in a fundamental region (of both duality groups) in moduli space.

In \cite{ad}, we compute the large $L$ asymptotics for
an ``index'' which counts supersymmetric vacua
with signs, and is thus a lower bound for the total number of supersymmetric
vacua.  The general formula for this is
\begin{eqnarray} \label{eq:dist}
I_{vac}(L \le L_{max})
 &=&  {(2\pi L)^{b_3}\over \pi^{n+1} b_3!}\int_{\CF\times\CH} 
\det(-R-\omega\cdot 1) ,
\end{eqnarray}
where $\CF$ is a fundamental region in the complex structure moduli space,
and $\CH$ is the fundamental region of $SL(2,\BZ)$ in dilaton-axion moduli
space.  This formula will have corrections in a series in $1/L$, but these
should be small if $L >> K$.

An important point about the result is that it gives not just the
number, but the actual distribution of points where 
the flux vacua stabilize the moduli.  The term $\det(-R-\omega\cdot 1)$ which
appears under the integral sign, is a $2n+2$ form, derived from
the K\"ahler form $\omega$, and the matrix of curvature two-forms $R$.
If we integrate this form over a subregion in moduli space, we obtain
the number of vacua which sit in this region.
These results can be
used to compare the numbers of vacua in different regimes 
of coupling and field space, as we will discuss in \cite{dd}.

While mathematical techniques exist to do this type of integral, 
the actual result is known at present only for
compactification on $T^6/\BZ_2$, for which
$K=b_3=20$ and $L=32$: one finds 
$$
I_{vac} = {7\cdot (2\pi L)^{20}\over 4\cdot 181440\cdot 12\cdot 20!} 
\sim 4\cdot 10^{21}.
$$

To complete the discussion, we need to discuss
stabilization of K\"ahler moduli.
Now no-scale structure is generically
spoiled by $\alpha'$ and non-perturbative corrections, say
$$
W_{NP} = e^{i N\rho} + \ldots .
$$
which, it has been suggested \cite{KKLT},
can be arranged to arise from some brane world-volume theory.
Very generally,
a solution of $DW(z)=0$ for the complex structure moduli,
with $e^K|W|^2 << M_{pl}^4$, will become a stable supersymmetric AdS
vacuum once these are taken into account.  \cite{KKLT,mrdstat}
For example,
$$
0 = D_\rho W = iNe^{iN\rho} - {3\over\rho-\bar\rho}W_{rest} 
$$
has a solution for
$$
{2N\over 3}(\Im\rho) e^{iN\rho} = W_{rest} .
$$
The function on the l.h.s. can take any value up to $2/3e \sim 1$, and
one expects an exact nonperturbative $W(\rho)$ to behave similarly.
Thus, any vacuum with $W_{rest}$ not too large, can be stabilized.\p

Thus, we need to know the distribution of AdS cosmological constants.
This distribution can be computed in the same way as above; details
will appear in \cite{dd}.  One finds that the distribution is ``uniform near
zero,'' in other words the number of vacua with small
$\Lambda=\epsilon << L$ goes as $N_{vac} \times \epsilon b_3/L$. 

There is a simple intuitive argument for this result.\cite{argument}
The AdS
cosmological constant can be regarded as the length squared of a
``vector in flux space'' defined by the values of the various periods
of cycles entering (\ref{eq:gvw}).  If we assume that this vector is
totally uncorrelated with the specific choice of flux, all vacua with
flux roughly orthogonal to this vector will have small cosmological
constant, leading to this result.  Now this assumption is not innocent
and is false in very similar examples (such as flux in the heterotic
string); however the actual computation in type \II\ does produce this
result.

Thus, taking $\epsilon\sim 10^{-3}$, we obtain a lower bound
$4\cdot 10^{18}$ on the number of flux vacua on $T^6/\BZ_2$.  
One might be even more
conservative and cut out the strong coupling regime, where the analysis
is not presently under control.  If we do this by
insisting on $\Im\tau > 40$, we find that $10^{17}$ 
vacua satisfy these constraints.\p

\section{Conclusions}

We discussed the statistical approach to string phenomenology, and
gave the first computation of the number of flux vacua in a specific
model.  This result confirmed the suggestion of Bousso and Polchinski
that this number should grow as $L^K$,
exponentially in the number of cycles $K$,
and is a first step in getting a solid idea of just how
many string theory vacua there are.

CY's are known with $K\sim 500$ and $L\sim 10^4$, so
there is some danger that the number of vacua is large enough to spoil
predictivity.  
We can roughly quantify the number at which we should start to worry,
as follows \cite{mrdstat}.  Consider the twenty dimensional parameter
space of the Standard Model couplings, along with the cosmological
constant.  Now, compute the volume in this space consistent with
present-day observations, where we measure each coupling in Planck
units (so, a coupling $\lambda$ of dimension $n$ has measure 
$M_{pl}^{-n} \int d\lambda$.  For the Standard Model, this comes out
to about $10^{-240}$, where $10^{-120}$ of this comes from matching
the cosmological constant, and the rest from the Higgs mass and other
couplings.

Now, suppose we consider string models which match the spectrum of the
Standard Model at low energies.
As we discussed, the fraction of four
dimensional models which do this is surely much greater than
$10^{-100}$, and $10^{-10}$ would seem to be a reasonable guess.

Take these models, and plot their couplings in this twenty dimensional
space.  The computation we discussed is also a first step towards
doing this, as it gives the distribution of vacua in moduli space, and
these moduli will control the couplings in the resulting low energy
theory.  To some approximation, the result (\ref{eq:dist}) says
that models are roughly uniformly distributed in moduli space; this
suggests that the dimensionless couplings could come out roughly
uniformly distributed as well.  This is less clear for dimensionful
couplings, and discussing these requires taking supersymmetry and its
breaking into account in more detail.  Let us continue however to
illustrate the idea, and return to this point.

Then, we can distinguish various cases.  If the number of models is
less than $10^{120}$, we do not expect to match the cosmological
constant, and need to find mechanisms which produce the observed small
value, to reasonably expect string theory to reproduce the data.  If
it is between $10^{120}$ and $10^{240}$, we expect models which
reproduce the cosmological constant to exist, and can in principle
test string theory by checking that the observed Standard Model
couplings are possible in some compactification -- just on statistical
grounds, this would be unlikely.

On the other hand, suppose there were many more than $10^{240}$
models, say $10^{1000}$ for definiteness.  In this case, one would
expect a vast number of models to reproduce the Standard Model in
every detail, just on statistical grounds.  If it furthermore turned
out that they led to many different predictions at higher energies,
one would clearly have grounds to worry that string theory was not
testable.

Of course supersymmetry changes these numbers.  With low scale breaking
(say $10 \TeV$), the corresponding ``volume in coupling space'' becomes
about $10^{-120}$.  Thus, unless the ratio of nonsupersymmetric
candidates to these candidates is greater than $10^{120}$, these
considerations might be regarded as favoring supersymmetry, as discussed
at more length in \cite{BDG}.  

There are some important points to be made here.  First, it is
essential to have some estimate of the {\it a priori} numbers, here
the ratio between the numbers of nonsupersymmetric and
supersymmetric vacua, to make this claim.  It is true that one does not
need a very accurate estimate of this ratio (here any accuracy better than
$10^{100}$ would suffice), but one does need some estimate, for this
argument to have any content beyond the standard argument
for supersymmetry from naturalness.

Second, it should be realized that ``favoring'' one mechanism over
another is interesting but not in itself decisive, if in fact both
types of vacuum exist.  We only observe one vacuum, and it might be of
either type.  On the other hand, suppose we estimated the number of
supersymmetric vacua which pass all tests as $10^{40}$, while the
number of non-supersymmetric vacua was estimated as $10^{-40}$.  This
could come out of the type of approximate estimates we are discussing,
and would mean that we need a coincidence which tunes parameters at
the $10^{40}$ level to get a candidate nonsupersymmetric vacuum.  If
our estimates were reliable, then the most reasonable interpretation
of such evidence would be that there are in fact no candidate
nonsupersymmetric vacua, and we would get a strong prediction from the
statistics of vacua.  It is this goal which motivates trying to get
controlled estimates for these numbers.

As a final comment, it is interesting that the numbers which are
coming out, of order $10^{100}$, are of the general order we need to
solve the cosmological constant problem, while not obviously spoiling
predictivity.  Thus, the picture we just outlined need not be
depressing -- there are many further conditions the correct vacuum
must satisfy, and it may be that further work along these lines will
demonstrate that only one or a few vacua work.  In any case, our main
point is that these are questions about string theory which, if the
theory has a precise definition, have definite answers, and that are
now becoming accessible to investigation.

\smallskip

I would like to thank S. Ashok, T. Banks,
F. Denef, S. Dimopoulos, M. Dine, S. Kachru, G. Kane, J. Maldacena,
G. Moore, B. Shiffman, S. Thomas, S. Trivedi, E. Witten and S. Zelditch
for collaboration and valuable discussions.

\end{document}